\documentstyle[12pt]{article}

\font\twlgot =eufm10 scaled \magstep1
\font\egtgot =eufm8
\font\sevgot =eufm7

\font\twlmsb =msbm10 scaled \magstep1
\font\egtmsb =msbm8
\font\sevmsb =msbm7

\newfam\gotfam

\textfont\gotfam\twlgot
\scriptfont\gotfam\egtgot
\scriptscriptfont\gotfam\sevgot

\newfam\msbfam
\textfont\msbfam\twlmsb
\scriptfont\msbfam\egtmsb
\scriptscriptfont\msbfam\sevmsb

\def\pBbb{\relax\ifmmode\expandafter\Bb\else\typeout{You cann't use
Bbb in text mode}\fi}
\def\Bb #1{{\fam\msbfam\relax#1}}

\def\thebibliography#1{\bigskip\section*{\centering
References\\}\bigskip\list
  {\arabic{enumi}.}{\settowidth\labelwidth{#1}\leftmargin\labelwidth
    \advance\leftmargin\labelsep
    \usecounter{enumi}}
    \def\newblock{\hskip .11em plus .33em minus .07em}
    \sloppy\clubpenalty4000\widowpenalty4000
    \sfcode`\.=1000\relax}

\newcommand{\ben}{\begin{eqnarray}}
\newcommand{\een}{\end{eqnarray}}
\newcommand{\be}{\begin{eqnarray*}}
\newcommand{\ee}{\end{eqnarray*}}
\newcommand{\bea}{\begin{eqalph}}
\newcommand{\eea}{\end{eqalph}}

\newcommand{\F}{{\bf F}}

\newcommand{\dr}{\partial}

\newcounter{eqalph}
\newcounter{equationa}

\newenvironment{proposition}[1]{\bigskip{{\sc Proposition} #1.}\it}{\bigskip}
\newenvironment{theorem}[1]{\bigskip{{\sc Theorem} #1.}\it}{\bigskip}
\newenvironment{proof}{{\bf Proof.}}{\bigskip}

\newenvironment{eqalph}{\stepcounter{equation}
\setcounter{equationa}{\value{equation}}
\setcounter{equation}{0}

\begin{eqnarray}}{\end{eqnarray}
\setcounter{equation}{\value{equationa}}}

\begin{document}
\hbox{}

\centerline{\bf \large GRAVITATION SINGULARITIES OF THE}
\medskip

\centerline{\bf \large CAUSTIC TYPE}
\bigskip

\centerline{\bf G.Sardanashvily}
\medskip

\centerline{Department of Theoretical Physics, Moscow State University}

\centerline{117234 Moscow, Russia}
\vskip1cm

Various criterion of gravitational singularities have been
suggested \cite{can2,fuc,haw}. In view of the well-known
correspondence between the space-time distributions and
the gravitational fields
on a world manifold $X^4$, gravitational singularities can be
indicated by singularities of these distributions \cite{sar,3sar}.
In the germ terms, singularities of a (3+1) distribution
look locally like singularities of a foliation whose leaves are level
surfaces of a real function  $f$ on $X$. If $f$ is a single-valued function,
changes of leave topology at critical points of $f$ take place
\cite{sar}. In case of a multi-valued function $f$,
one can lift the foliation to the total space of
the cotangent $t^*X$ bundle over $X$, then extend it over branch points
of $f$ and project this extension onto $X$. Singular points of this
projection constitute a Lagrange map caustic by Arnol'd.
\bigskip

In  gravitational  theory, a space-time structure is usually defined to be
a (3+1) decomposition
\begin{equation}
TX = \F\oplus T^0X \label{2.12}
\end{equation}
of the tangent bundle over a world manifold into  a  3-dimensional
spatial  subbundle   $\F$   and its time-like orthocomplement $T^0X$.

Recall that there is the 1:1 correspondence  between the
nonvanishing 1-forms $\omega$ on a manifold $X$ and the smooth  orientable
distributions $\F$ of 1-codimensional subspaces of tangent spaces to $X$.
This correspondence is defined by the equation
\[\F\rfloor \omega = 0.\]
A form $\omega$ is called a generating form\index{generating form} of a
distribution \F.

We have the following theorem \cite{haw,sul}
formulated in terms of space-time distributions \cite{3sar}.

\begin{theorem}
For every gravitational field $g$ on a world  manifold
$X$, there exists an associated pair $(\F,g^R)$ of  a
space-time distribution $\F$ with a generating  tetrad  form
\[
h^0=h^0_\mu dx^\mu
\]
and  a Riemannian metric $g$, so that
\begin{equation}
g^R=2h^0\otimes h^0-g = h^0\otimes h^0  + k \label{2.16}
\end{equation}
where $k$ is a Riemannian metric in the tangent subbundle $\F$. Conversely,
given a Riemannian metric $g^R$,  every  oriented  smooth  3-dimensional
distribution $\F$ with a generating form $\omega$ is a  space-time
distribution compatible with the gravitational field $g$ given  by
expression~(\ref{2.16}) where
\[h^0=\omega/\mid\omega\mid,\qquad \mid\omega\mid^2=g^R(\omega,\omega) =
g(\omega,\omega). \]
\end{theorem}

A 1-codimensional distribution $\F$ is called an integrable
distribution if its generating form
$\omega$ obeys the equation
\[\omega\wedge d\omega = 0.\]
In this case, fibres of the corresponding (3+1) decomposition
(\ref{2.12}) are
tangent to leaves of some 1-codimensional foliation
of spatial hypersurfaces of a world manifold $X$. An
integrable space-time
distribution (a space-time foliation) is
called causal if its generating form can be exact, that is,
\[\omega = df\]
where $f$ is some real function  on  $X$  which has no  critical
points where $df=0$ \cite{law}.
This notion of causality  coincides  with  the
one of stable causality by Hawking \cite{haw}. Leaves of a  causal
foliation are level surfaces
of its generating function $f$. No curve
transversal to leaves of a causal foliation intersects  each leave more
than  once.

We say that a gravitational field $g$ on
$X$ is free from singularities if there exists an associated pair of a
complete Riemannian metric $g^R$ and a causal space-time foliation
{\bf F} with the generating form $h^0$ such that $g^R(\nabla h^0, \nabla
h^0)$ is bounded on $X$. This condition guaranties that, being complete
with respect to $g^R$, a space-time satisfies the well-known
$b$-completeness condition \cite{can}.

One can distinguish several types of gravitation singularities in
accordance to this criterion. We examine gravitation
singularities characterized by singularities of space-time distributions.
The distribution singularities can be described locally (in the germ
form) as singularities of a causal foliation with a generating
function $f$. There are two types of these singularities.

(i) A single-valued generating function $f$
has critical points where $df=0$.
It generates the Haefliger structure (the singular foliation)
of its level sets on $X$. These level sets
change their topology at critical points
of $f$. Gravitation singularities of this type are the
scalar curvature singularities in accordance to the classification in
\cite{ell}.

(ii)
A generating function $f$ is a multiple-valued function on $X$. The
leaves of the foliation $\F$ defined on the domain where $f$
is a single-valued begin to
intersect each other at branch points of $f$. Branch points of $f$ where
the foliation is destroyed form a caustic. To describe foliation
singularities of this type, one can lift the space-time foliation $\F$ into
the total space of the cotangent  bundle $T^*X$, then extend this
lifted foliation over the singular points, and project
the extended foliation
onto the base $X$. Singularities of $\F$ can be described as
singularities of this projection.

In gravitation theory, a geometrical locus of focal and conjugate points
is called a caustic by analogy with geometrical optics \cite{ros,war}. We
follow the general mathematical notion of caustics as singularities of
the Lagrange maps \cite{arn,fri}. Each
caustic can be brought locally (in germ
terms) into the following standard form.

Let a space ${\bf R}^{2n}$ be endowed with the coordinates
$\{x^\mu,P_\mu\}$. Let us consider the Liouville form
\begin{equation}
\alpha=P_\mu dx^\mu \label{2.20}
\end{equation}
on ${\bf R}^{2n}$ and a submanifold $N$ of ${\bf R}^{2n}$ such that
\[d\alpha(N)=0,\]
that is, being restricted onto $N$, the form $\alpha$ is exact:
\[\alpha(N)=dz(N).\]
Such a manifold of maximal dimension $n$ is called a
Lagrange submanifold.  A Lagrange
submanifold can be defined by a generating function $S(x^i,P_j)$ of $n$
variables $(x^i,P_j, i\in I, j\in J)$ (where $(I,J)$ is some
partition of the set $(1,...,n)$. It is given by the
relations
\[x^j=-\frac{\dr S}{\dr P_j},\qquad P_i=\frac{\dr S}{\dr x^i}.\]

Let us consider the projection
\[\pi: (x^\mu,P_\mu)\to (x^\mu)\]
of ${\bf R}^{2n}$ onto ${\bf R}^{n}$. Being
restricted to the Lagrange submanifold
\[\pi_N: (x^i,P_j)\to (x^i,x^j=-\frac{\dr S}{\dr P_j}),\]
this projection is called the Lagrange map. A
caustic is defined to be the set of critical points of a
Lagrange map, i.e., the points where the matrix
\[\dr^2S/\dr P_i\dr P_j\]
is singular.

For instance, a caustic on manifolds is defined as follows. Let the
cotangent bundle $T^*X$ be provided with the induced coordinates
$(x^\mu,P_\mu,...x_\mu)$. The Liouville form (\ref{2.20}) defines
$n$-dimensional Lagrange submanifolds of $T^*X$. Singular points of
projection of such a Lagrange submanifold onto the base $X$ form
a caustic.

Let us note that a geometrical locus of
focal and conjugate points of Riemannian
and time-like pseudo-Riemannian geodesics also is a caustic in accordance
to the Arnol'd definition \cite{sar}.

Our definition of foliation caustics is based
on the following proposition.

\begin{proposition}{}
For any foliation of level surfaces $\F$ of a manifold $X$, there is a
foliation $\F'$ of some Lagrange submanifold of $T^*X$ such that $\F$
is the image of $\F'$ under the Lagrange map.
\end{proposition}

\begin{proof} Let $f$ be a generating function of the
foliation $\F$.  We define the embedding
\[
\gamma: (x^\mu)\to (x^\mu,P_\mu=\frac{\dr f}{\dr x^\mu})
\]
of $X$ into $T^*X$. Its image is a Lagrange submanifold of $T^*(X)$. Let
$\F'$ be the induced foliation $\pi^*_{\gamma(X)}\F$ of
$\gamma(X)$ where $\pi_{\gamma(X)}$ is
the Lagrange map
\[\pi_{\gamma(X)}: \gamma(X)\to X.
\]
Since $\gamma$ and $\pi_{\gamma(X)}$ are diffeomorphisms between $X$
and $\gamma(X)$ ($\pi_{\gamma(X)}\circ\gamma={\rm id} X$), the foliation
$\F$ on $X$ can be represented as the image of the foliation $\F'$ on
$\gamma(X)$ under the Lagrange map $\pi_{\gamma(X)}$.
\end{proof}

For instance, let $N\subset T^*X$ be the Lagrange submanifold
generated locally by a function $S(x^i,P_j)$, and let $\F'$ be the
foliation of level surfaces of the function
\[f'(x^i,P_j)=S-P_j\frac{\dr S}{\dr P_j}\]
on the Lagrange submanifold $N$. The image $\pi_N(\F')$ of $\F'$ under the
Lagrange map $\pi_N$ is a foliation of the image $\pi_N(U)$ of the domain
$U\subset N$ where this Lagrange map has no critical points. This
foliation is destroyed at caustic points of the Lagrange map $\pi_N$.

There are the following classes $A_2, A_3, A_4, D_4, A_5, D_5$ of stable
caustics on a 4 dimensional manifold. For example, the canonical
generating function of caustic $A_3$ takes the form
\[S=-P^4_0+x^1P_0^2,\]
and the corresponding Lagrange manifold $N$ is given by equations
\[x^0=4P_0^3-2x^1P_0,\qquad P_1=P_0^2.\]
The Lagrange map then reads
\begin{equation}
x^0=4P^3_0-2x^1P_0. \label{2.21}
\end{equation}
The caustic set where
\[\dr^2S/\dr P_0^2=0\]
consists of the points
\[x^1=6P_0^2,\]
and its Lagrange image on $X$ contains the points
\[(x^0)^2=\frac{8}{27}(x^1)^5.\]
The generating function of the foliation $\F'$ on the Lagrange manifold
takes the form
\[f'=3P_0^4-x^1P_0^2.\]
Then, on $X$, the generating function of the Lagrange image of $\F'$
reads
\[f(x^0,x^1)=f'(x^1,P_0(x^0,x^1))\]
where the function $P_0(x^0,x^1)$ is defined by equation (\ref{2.21}). The
function $P_0$ and the generating function $f$ become
three-valued functions at caustic points. The $A_3$-germ of foliation
caustics thus is characterized by the behavior of the component
$\omega_0$ of the foliation generating form which is tripled at caustic
points.

Caustic singularities have the following feature. There are domains of
a space-time where not nearest, but the far separated leaves begin to
intersect each other. Therefore, a space-time foliation can be
locally prolonged over the caustic points, whereas global prolongation of
this foliation is impossible.

For example, let $f(u,v)$ be a real function on ${\bf R}^2$ which obeys
the equation
\[f^3(u,v)-3uf(u,v)-2v=0\]
where $u,v$ are coordinates on ${\bf R}^2$. This function is the
singled-valued one
\[f_+=[v+(v^2-u^3)^{1/2}]^{1/3}+[v-(v^2-u^3)^{1/2}]^{1/3}\]
on the domain $U=(u,v: v^2>u^3)$, and it is the three-valued
function
\be
&&f_{0,1,2}=2u^{1/2}{\rm cos}(\frac13(\phi+2\pi n)),\\
&&\phi={\rm arccos}(vu^{-3/2}),\qquad n=0,1,2,
\ee
at points $v^2<u^3$. Let $\F$ be the foliation
\[\F_c=\{u,v: f_+(u,v)=c={\rm const}\}\]
on $U\subset {\bf R}^2$. Its leaves $\F_c$ are the lines
\[2v=c^3-3uc,\qquad -\infty<c<+\infty.\]
This foliation has the caustic singularity at the branch points $v^2=u^3$ of
the function $f$. Moreover, $u=v=0$ is the $A_3$-caustic point, whereas
the other ones $v^2=u^3\neq 0$ are points of the $A_2$-caustic. The leaves
$\F_c, c>\alpha>0$, can be prolonged over the caustic curve $v=u^{3/2}$
onto the domain $0<v<u^{3/2}$ where they can be described as leaves of the
foliation
\[f_0(u,v)={\rm const}.\]
These leaves however begin to intersect
each other when $v<0$, although the nearest leaves intersect each other
only on the caustic curve $v=-u^{3/2}$. Note that the leaves $\F_{c>0}$ begin
to intersect the leaves $\F_{c<0}$ on the caustic curve $v=u^{3/2}$.

Caustic singularities however are not reduced to the locally extensible
singularities \cite{ell}. For instance, the $A_2$-caustic points
$u^2=v^3\neq 0$ of the above-mentioned foliation caustic are locally
extensible singularity points, whereas the $A_3$-caustic point $u=v=0$ is
not locally extensible.

\end{document}